\RequirePackage{fix-cm}
\documentclass{svjour3}                     % onecolumn (standard format)
\smartqed  % flush right qed marks, e.g. at end of proof
\usepackage{graphicx}
\usepackage{epsfig}
\usepackage{epstopdf}
\usepackage{parskip}
\usepackage{color}
\usepackage{hyperref}
\usepackage{lineno}
\usepackage{cite}
\usepackage{url}
\usepackage{amsmath}

\begin{document}

\title{Approximate Solution Approach and Performability Evaluation of Large Scale Beowulf Clusters}



\author{ Yonal Kirsal  \and
        Yoney Kirsal Ever
     }


\institute{1 \at
              Electrical and Electronic Engineering Department, European University of Lefke, Lefke, North Cyprus \\
                            \email{ykirsal@eul.edu.tr}           %  \\
           \and
           2 \at
              Software Engineering Department, Near East University, Nicosia, Mersin 10, Turkey \\
              \email{yoneykirsal.ever@metu.edu.tr}
           }


\maketitle

\begin{abstract}

Abstract Beowulf clusters are very popular and deployed worldwide in sup- port of scientific computing, because of the high computational power and performance. However, they also pose several challenges, and yet they need to provide high availability. The practical large-scale Beowulf clusters result in unpredictable, fault-tolerant, often detrimental outcomes. Successful development of high performance in storing and processing huge amounts of data in
large-scale clusters necessitates accurate quality of service (QoS) evaluation. This leads to develop as well as design, analytical models to understand and predict of complex system behaviour in order to ensure availability of large- scale systems. Exact modelling of such clusters is not feasible due to the nature of the large scale nodes and the diversity of user requests. An analytical model for QoS of large-scale server farms and solution approaches are necessary. In this paper, analytical modelling of large-scale Beowulf clusters is considered together with availability issues. A generic and flexible approximate solution approach is developed to handle large number of nodes for performability evaluation. The proposed analytical model and the approximate solution approach provide flexibility to evaluate the QoS measurements for such systems. In order to show the efficacy and the accuracy of the proposed approach, the results obtained from the analytical model are validated with the results obtained from the discrete event simulations.

\keywords{Beowulf clusters \and approximate solution technique \and analytical modelling\and performability evaluation \and large-scale clusters \and high performance computing}
\end{abstract}

\section{Introduction}
\label{intro}

The rapid development of computers, parallel and distributed systems recently provide flexible, efficient, and highly available services to their users. Thus, users increasingly expect better quality of service (QoS) from such systems.

Due to high QoS expectation, the computer/communication system and cloud- based networks have to be highly available and being easily managed [1].
In order to provide the best QoS and the seamless service to the users various kinds of computing paradigms being effectively used. Parallel computing is most commonly used approach in such systems. It is a way of dividing a large tasks into smaller tasks and using more than one node simultaneously to perform these divided tasks [2]. Hence, parallel processing provides high performance with reducing the time by sharing the necessary work among the cluster nodes to perform complex computations [3]. In addition, high- performance computing (HPC) uses of supercomputers and parallel process- ing techniques for solving complex computational problems [4?6]. HPC delivers sustained performance by offering storage resources. However, computations for parallel processing are very expensive and because of this reason many organizations cannot afford to have them. This leads to the introduction different architectures such as symmetric multiprocessors, vector processors and cluster computing.

The Beowulf clusters have been widely used for clustering. Beowulf clusters allow large amount of computing nodes to be available for parallel or simultaneous processing at much lower cost [7?9]. Beowulf clusters are scalable performance clusters based on a multi-processor system that consist of commodity hardware on a private system network with open source software infrastructure [10]. In a Beowulf cluster, network of computers is tightly connected that are dedicated to simultaneously provide service to incoming task requests. The typical Beowulf cluster generally has two types of nodes: A head node (master node) and the identical computing nodes [11]. The head node main duty is serving and distributing the user requests to computing nodes. Computing nodes are usually dedicated to service. In addition, the computing nodes can- not serve the tasks if the head node is not operative [12]. Due to the single head node formation, such systems are vulnerable. The head node failures affect the availability of the clusters significantly and therefore, this limits access to the healthy identical nodes. Additionally, depending on the structure of the cluster, head node may or may not be part of the computation.

In this paper, performability models have been formed which takes both performance and availability concerns into account for large scale Beowulf clusters an analytical point of view. It is essential to evaluate the performability out- put parameters in order to obtain QoS of the system. Thus, the availability of multi-nodes systems can be affected by nodes failure due to various reasons. The single head node architecture makes the performability analysis even more interesting and essential. Therefore, in order to obtain more realistic QoS measurements and analysis of Beowulf clusters, performance models and availability should be employed together for such server farms. Although, analytical approaches are presented for server farms (with one head and several identical computing nodes) as well as Beowulf clusters, large-scale systems could not be considered due to the state space explosion problem [13]. Existing modelling and exact solution approaches such as the spectral expansion method is not able to handle such large networks [14]. In [15] a situation of the state space explosion problem is encountered in a spectral expansion method where Beowulf clusters are modelled and solved for various performance measures. Thus, the number of parallel servers considered could not exceed due to state space explosion. However, cloud computing, web server and cluster computing provide a total of 256, 372 or even more (i.e., 512) nodes [16,17]. The well-known exact solution approaches to availability issues or open queuing networks with failure suffer from similar problems. In addition, solving complex systems through simulation resulted in prohibitively long computational times for such systems. Therefore, it is of great importance to develop an analytical method and solution approach to overcome these problems for such systems. The main contribution of the paper can be summarized as follows:
\begin{itemize}
\item {An approximate and flexible three dimensional (3D) solution approach is presented in large-scale Beowulf clusters for performability evaluation. Al- though the Markov models given are not new, the approach is new and it can be used for large-scale Beowulf systems where large numbers of computing nodes can be considered typically up to several hundreds or thousands.
}
\end{itemize}

The performability results obtained from the analytical model are compared to the discrete event simulation (DES) results in order to show the accuracy and effectiveness of the proposed work. Findings show that the analytical modelling and an approximate solution approach presented in large-scale Beowulf clusters providing a high degree of accuracy, and it is significantly more efficient than the simulation approach in terms of computational time.

The rest of the paper is organized as follows: Section 2 presents the related studies and gives the motivation of the study. Section 3 describes the system model and the analytical solution approach. Section 4 presents the numerical results and discussions for the proposed model. Finally, Section 5 concludes this paper.

\section{Related Works}
The Beowulf clusters have been widely used all over the world. They can be employed for coarse grained applications such as Monte Carlo calculations, statistical simulations, high throughput applications, and also can be used in grid computing. Beowulf type cluster system is a good example of HPC systems. This is due to a single head node which is possible to have a backup node for the head node. These types of systems are called highly available Beowulf systems [18,19].

HPC is the future paradigm that has been dominating the visualization and processing of huge amount of web data. HPC systems fundamentally provide access to large pools of data and computational resources through a variety of interfaces similar to the existing grid, HPC resource management and programming systems [20]. The HPC service providers must strive to ensure good QoS by offering highly available services with dynamically scalable resources as stated in [19]. In [19] authors used HA-OSCAR, which is an open source High Availability (HA) solution for HPC/cloud that offers component redundancy, failure detection, and automatic failover. It is assumed that any task sent to the cloud/cluster center is serviced with a suitable node which is called a facility node [21]. When the task is serviced, it leaves the center. This facility node may contain different computing resources such as web servers, database servers, directory servers, and others. However, in [21] emphasized that cloud centers differ from traditional queuing systems in a number of important aspects. However, more importantly, a cloud center can have a large number of orders of hundreds or thousands of facility nodes that, traditional queuing analysis rarely considers systems of this size. In 2006, Amazon introduced the elastic computing cloud (EC2) that customers could rent, by the hour, Xen- based virtual machines hosted in Amazon?s data center [22]. In this, users have full root-level access to virtual machines so that they can fully customize and optionally publish machine images. However, in [23], authors subordinated computer nodes. These nodes are configured from the head node. Additionally, the authors used the Rocks toolkit. The Rocks toolkit is such a methodology and more than 2000 clusters have been built with open-source software stack.

Homogeneous multi-server systems have been considered with different repair strategies for performability evaluation in the literature [5, 7, 10-12, 14, 15, 27,29,31]. However, the proposed models and solution approaches used are more applicable to small and/or medium size systems rather than large-scale systems. When the large-scale systems are considered, the state space explosion is a general problem for state space representation of queuing systems. The state space explosion is a general problem for state space representation of queuing systems, but it is encountered especially in multiprocessor systems. The spectral expansion and the matrix geometry method solve this problem partially. They are capable to consider systems with infinite queuing capacities. However, when the number of servers increase rapidly they face with analytical difficulties. Large scale Beowulf clusters face the space explosion problem as well. In [7] and [15] the Beowulf clusters are modelled and solved for various measures, but the identical computing nodes considered could not exceed some limitations due to the state space explosion problem. Thus, in this paper, the proposed model aim is to solve this problem for a large number of nodes in Beowulf clusters. The given approach is new and flexible. Thus, it can be applicable to similar large scale and complex systems.

As stated in [24], the majority of current cloud computing infrastructure con- sists of services that are offered up and delivered through a service center such as a data center, that can be accessed from a web browser anywhere in the world. Our proposal also relies on that. As the population size of a typical cloud center is relatively high while the probability that a given user will request service is relatively small, the arrival process can be modelled as a Markovian process [25]. In [26] an open Jackson queueing network based model is used to characterize the service components in content-delivery-as-a- service (CoDaaS). A Jackson network is constructed with a network of queues, where the arrivals at each queue are modelled as a Poisson process, and the service times follow the exponential distribution. In [27] queuing theory is used to identify and manage the users? response time for services. In [28], the authors obtained the response time distribution of a cloud system modelled on a classic $M/M/m$ open network, assuming an exponential density function of the inter-arrival and service times. In [29], the authors obtained the response time distribution for a cloud with an $M/M/m/m+r$ system model. Both inter- arrival and service distribution times were assumed to be exponential and the system had a finite number of $m+r$ size buffers. In [30], a queuing performance model consisting of a cloud architecture and a service center such as a data center is studied. The service center is taken as a database server. This means that both the time between user arrivals to the system and the service time of the system follow an exponential distribution with means $\lambda$ and $\mu$ respectively, with m servers with a first come first serve (FCFS) scheduling policy in [30].

In [31] the availability modelling and evaluation of HPC computing systems is presented. The necessity of availability is also emphasized and shown. The authors in [31] have developed a novel solution approach using an object oriented Markov model which provides availability modelling for typical high- performance cluster computing systems. Numerical results presented in [31] demonstrated that availability modelling and evaluation need to be considered at the system design stage for typical high-performance cluster computing systems.

\section{The Proposed System and The Approximate Solution Approach}

In this section, a homogeneous multi-nodes system is presented for performability evaluation of large-scale Beowulf clusters with failures and repairs. As defined before, a typical Beowulf cluster has two types of nodes: A head node and multiple identical computing nodes as shown in Fig. 1. The head node may or may not serve the task, however, the main responsibility is distributing tasks to computing nodes. Identical computing nodes normally provide computation. In this paper, the head node does participate to computations.

 \begin{figure}[ht!]
\begin{center}
      \includegraphics[height=3.0in]{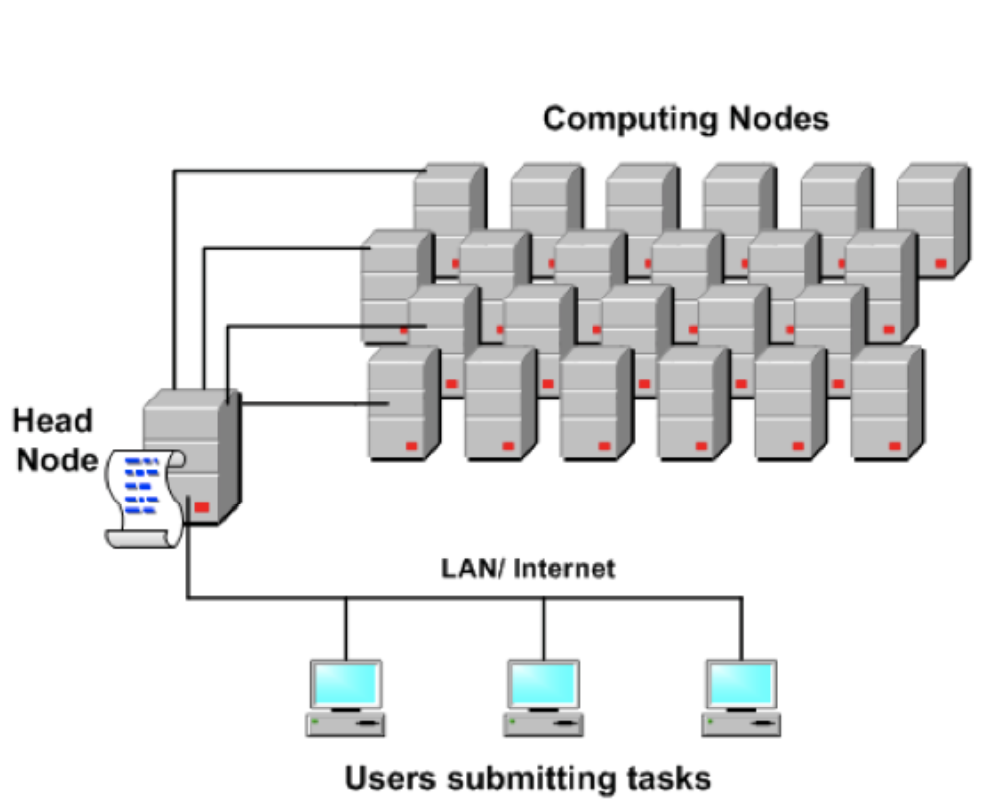}
      \caption{Multi-nodes Beowulf cluster architecture}
      \end{center}
\end{figure}

If the head node is not working, the computing nodes cannot serve the tasks. This is because of the head node is responsible for the organization and the distribution of tasks. Due to the head node failure, identical computing nodes are vulnerable. The failure of head node limits access to healthy identical computing nodes. The Beowulf multi-nodes system is shown in Fig. 2. The system consists of a head node $(1)$ and $S-1$ identical parallel computing nodes, numbered $2, 3,. . ., S,$ with a bounded common queue. $L$ is the queue capacity of the proposed system where $L\geq S$. Tasks arrive at the system in a Poisson stream at a mean rate of $\lambda$, and join the queue. Tasks are homogeneous and the service rates of the identical computing nodes are equal. The service times of requests serviced by the computing node $k (k=2,. . .,S)$ and the head node are distributed exponentially with mean $1/\mu$ and $1/\mu h$, respectively. Even though, if the head node participates in computations, it generally has the same service rate as that of the identical computing nodes $(\mu=\mu h)$.

\begin{figure}[!htbp]
  \begin{center}
    \leavevmode
    \ifpdf
      \includegraphics[height=2.5in]{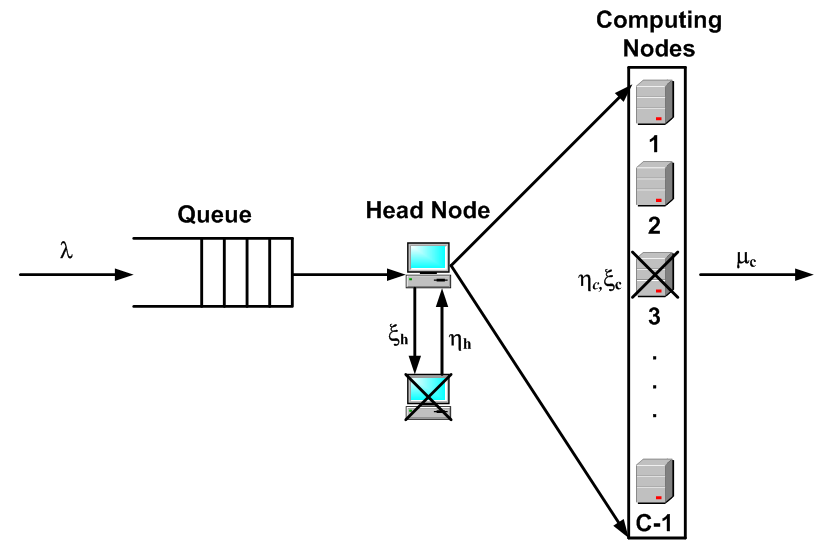}
              \fi    \caption{The proposed system considered with failures and repairs }
    \label{SM}
  \end{center}
\end{figure}

$1/\xi h$ and $1/\xi$ are operative periods of head node and computing nodes, respectively, and the means are also distributed exponentially. Thus, $\xi h$ and $\xi$ are failure rates of the head node and the identical computing nodes, respectively. At the end of the node $k (k=2,..., S)$ failure time an exponentially distributed repair time is needed with mean $1/\eta$. On the other hand, if head node fails, the repair rate is provided with mean repair time 1/?h. The repair priority is given to the head node when the head node and more than one computing nodes fail at the same time. This is because, the identical computing nodes cannot serve without of the head node. If there are requests waiting to be served, the operative computing nodes cannot be idle. In addition, the computing nodes serve with higher service rates, if the number of operative computing nodes is more than the number of requests in the system. Services that are interrupted by fails are eventually resumed from the point of interruption or repeated with re-sampling. In case of head node failures tasks continue to arrive with the same rate, $\lambda$ and, tasks in the queue remain in the queue without being serviced.

\subsubsection{Modelling Proposed Beowulf Clusters}
In this section, the proposed analytical model and approximate solution methods are introduced for large-scale Beowulf clusters. It is possible to represent the propose system with $S$ computing nodes, including the head node, by using a Quasi Birth and Death (QBD) process with finite state space. Since in Beowulf systems, none of the computing nodes can operate without the head node, the relation of the failure and the repair rate of the head node leads us to model the proposed system in two phases. The Fig. 3 indicates relationships between two phases. The first phase is used in states where the head node is always available which is indicated as $Plane_1$. The $Plane_0$ is the second phase, which, is used to represent the states where the head node is broken. Hence, the proposed system has two phases as shown in Fig. 3. Thus, the proposed system can be presented in three dimensions (3D).

\begin{figure}[!htbp]
  \begin{center}
    \leavevmode
    \ifpdf
      \includegraphics[height=1.5in]{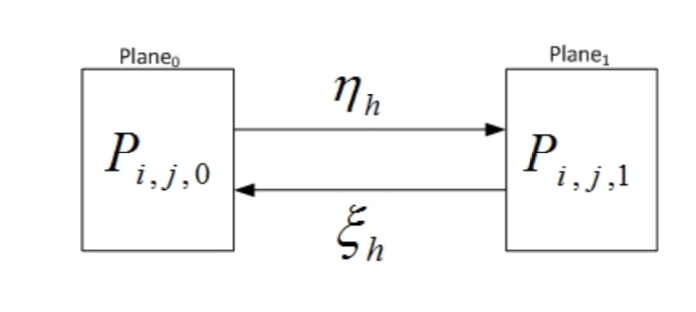}
              \fi    \caption{General transitions between two phases}
    \label{SM}
  \end{center}
\end{figure}

$P_{i,j,n}$ are all the steady state probabilities of the proposed system. It can be seen in the Fig. 3, value of $i$ and $j$ indicate number of computing nodes and number of tasks in the system, respectively. $n$ indicates the mode of a head node. When $n=0$ the head node does not operate and it represents the Plane0. On the other hand, when $n=1$ the head node is operative and these states are represented in $Plane_1$.
Figs. 4 and 5 show the state diagram of the $Plane_0$ and $Plane_1$, respectively of the proposed 3-D system. There are $S-1$ computing node configurations, $i=0, 1,..., S-1$ in Fig. 4. These $S-1$ configurations indicate the possible states of the $Plane_0$. $L$ represents the number of tasks/requests in the system for both figures. The $Plane_0$ describes case of the head node is not operative and the whole system does not provide services as shown in Fig. 4. Therefore, the downward transitions with service rate $\mu$ are not available. 

\begin{figure}[!htbp]
  \begin{center}
    \leavevmode
    \ifpdf
      \includegraphics[height=2.5in]{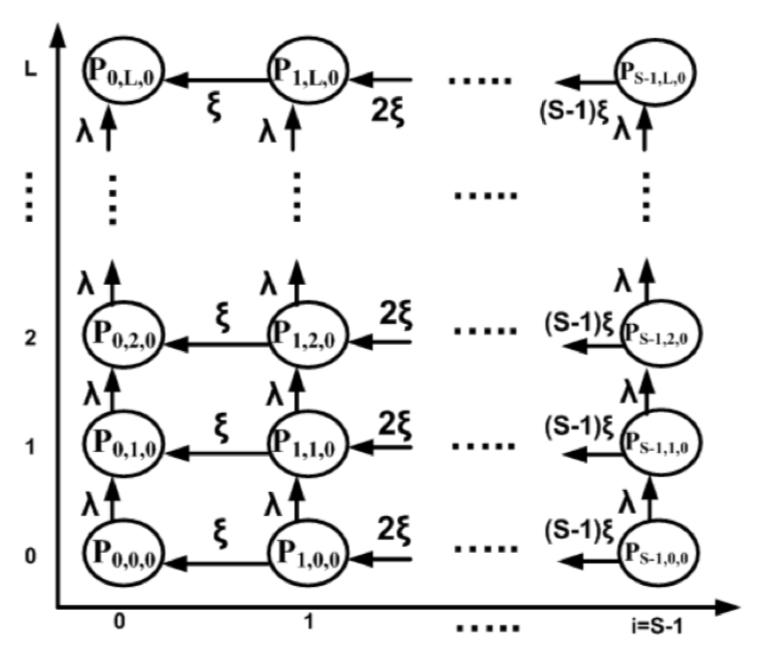}
              \fi    \caption{State diagram for the $Plane_0$ where head node is not working}
    \label{SM}
  \end{center}
\end{figure}

The repair priority is given to head node and the identical computing nodes cannot be repaired before the head node became alive. In other words, there are no repair transitions for the identical computing nodes in $Plane_0$, since the only repair transition which can take place is the transition to $Plane_1$. On the other hand, in Fig. 5 there are $S$ computing node configurations, $i=1,..., S$ and they are used to represent possible operative sates similar to Fig. 4. However, the number of computing nodes starts from one because the head node is operative. Downward transitions are possible since the system is alive. It is possible to use the repair facility in order to deal with identical computing node failures since the head node is operative. Available identical computing nodes can provide service with a service rate of $\mu$ and the broken nodes can be repaired with rate of $\eta$.

\begin{figure}[!htbp]
  \begin{center}
    \leavevmode
    \ifpdf
      \includegraphics[height=2.5in]{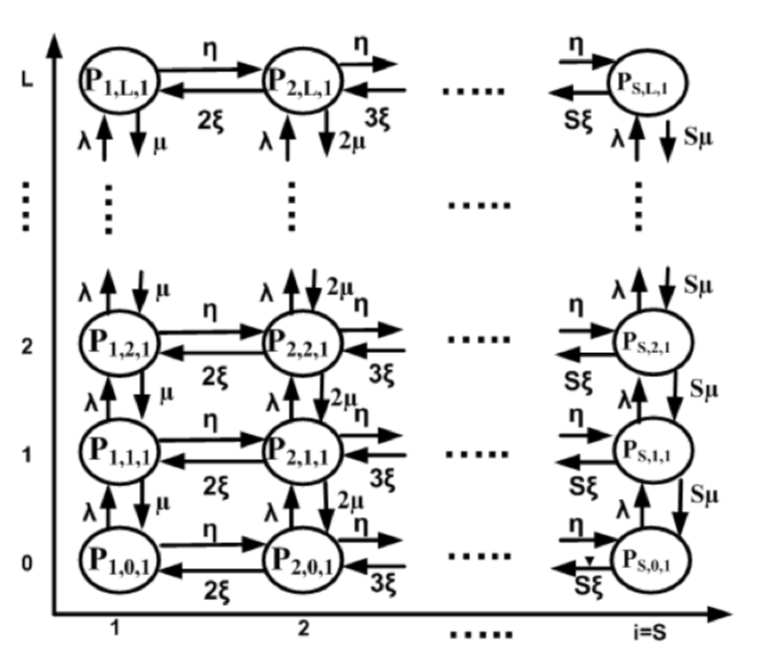}
              \fi    \caption{State diagram for the $Plane_1$ where head node is working}
    \label{SM}
  \end{center}
\end{figure}

\subsection{The Approximate Solution Approach}

This section explains and represents the proposed approximate solution approach for large-scale Beowulf clusters with availability issues. The analytical evaluation is considered and relevant equations are generated. An approximate decomposition is applied to implement to Beowulf multi-nodes cluster in order to fast convergence of the steady state probabilities. Then, all balance equations are produced considering the proposed system for a large number of parallel homogeneous computing nodes of the Beowulf cluster. The mathematical equations are produced to get an approximate solution for the performability of large scale Beowulf clusters. All required performance measurements can be obtained by generating the balance equations for steady state probabilities of the system and solving them by using an iterative method for both planes. Moreover, in order to have a faster convergence of iterative solution, the state probabilities of $P_{i,j,1}$ can be computed by analytical decomposition. These state probabilities of $P_{i,j,1}$ may not give very close approximations to real steady state probabilities however the balance equations are still required in order to take all possible transitions into account for fast convergence.
Therefore, mathematical equations are required in order to have an approximate solution for these probabilities. Thus, to find all $P_{i,j,n}$, the sum of all probabilities in both planes of the computing nodes should be considered individually. Two planes, $Plane_0$ and $Plane_1$, can then be analyzed separately. Every single plane has its own states and sum of all these state probabilities are not equal to one like single/multi server queue system. Thus, it is necessary to compute the sum of all probabilities in each phase. The sum of the overall probabilities $(Plane_0 + Plane_1)$ should be one. In order to obtain a general solution for the sum of overall probabilities in each plane, the equation 1 can be used.
\begin{equation}
P_{i,j,0} + P_{i,j,1} = 1
\end{equation}

Therefore, the equations 2 and 3 are derived for $Plane_0$ and $Plane_1$, respectively.
\begin{equation}
P_{i,j,0} = \frac{\xi h}{ \eta h + \xi h}
\end{equation}

\begin{equation}
P_{i,j,1} = \frac{\eta h}{ \eta h + \xi h}
\end{equation}

Both equations clearly indicate that the head node failure and repair rates are essential to find overall probabilities for such systems considered. Therefore, the following actions are taken to analyze performability of the Beowulf multi-node systems. The state probabilities of $Plane_1$ has been analyzed where the head node is operative. Hence, the sum of all possible probabilities of $Plane_1$ is taken as $\eta h/ \eta h+ \xi h$ as it can be derived from equation 3 above. Hence, equation 4 can be written as follows:

\begin{equation}
\sum_{i=1}^{S}\sum_{j=0}^{L} P_{i,j,1} = \frac{\eta h}{\eta h + \xi h}
\end{equation}

It is required to obtain the $P_{i,j,1}$ values in $Plane_1$. However, it is clear that
these probabilities cannot be obtained directly by using a product form solution. Therefore, first the sums of all probabilities are considered for each operative state of the system in $Plane_1$. Figure 6 shows the overall operative states of computing nodes for $Plane_1$.

\begin{figure}[!htbp]
  \begin{center}
    \leavevmode
    \ifpdf
      \includegraphics[height=0.7in]{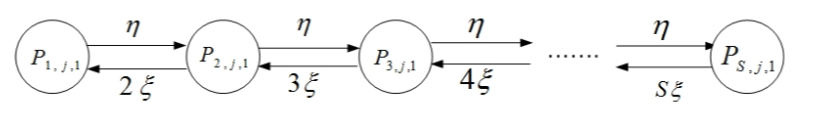}
              \fi    \caption{General lateral transitions of the system for a $Plane_1$}
    \label{SM}
  \end{center}
\end{figure}

Equation 5 can then be used to derive to calculate $P_{i,j,1}$ for all possible values of $i$.

\begin{equation}
\sum_{j=0}^{L}P_{i,j,1} = {\frac{\frac{\eta h}{\eta h + \xi h}}{\sum_{j=0}^{S} \frac{1}{i!}(\frac{\eta}{\xi})^{i-1}}}{\frac{1}{i!}(\frac{\eta}{\xi})^{i-1}}
\end{equation}

where $i=1,2,3, . . . S$. Since the sum of all $P_{i,j,1}$ in $Plane_1$ is known, it is easy to compute the overall probabilities for each operative computing node. $P_{i,j,1}$ can then be calculated in terms of $P_{i,0,1}$. Hence, equation 6 is derived using product form solution [32]. By these set of equations all $P_{i,j,1}$ can then be expressed in terms of $P_{i,0,1}$.

\begin{equation}
P_{i,j,1} = \begin{cases}
      \frac{\rho^i}{j!}\cdot P_{i,0,1}, &      \\
      & \\
      \frac{\rho^i}{j! i!^{i-1}}\cdot P_{i,0,1}, & i+1 \leq j \leq L 
    \end{cases}
\end{equation}

where $\rho = \lambda/\mu $and $1 \leq i \leq S$. Then, equation 6 can be generalized for each column as follows:

\begin{equation}
\sum_{j=0}^{L}P_{i,j,1} = \Bigg [ \sum_{j=0}^{i} \frac{\rho^i}{j!} + \sum_{j=i+1}^{L} \frac{\rho^i}{i!i^{j-i}} \Bigg ] P_{i,0,1}
\end{equation}

where $i=0,1,. . . , S$. Hence, $P_{i,0,1}$ can be computed as in equation 8 using the equations 5-7 with some simplifications.

\begin{equation}
P_{i,0,1}= \frac{(\frac{\eta}{\xi})^i}{i!\sum_{k=0}^{S}{\frac{(\frac{\eta}{\xi})^k}{k!}}}\Bigg [ \sum_{j=0}^{i} \frac{\rho^i}{j!} + \frac{(i^{L-i}\rho^{i+1})-\rho^{L+1}}{i!i^{L-1}(i-\rho)} \Bigg ] ^{-1}
\end{equation}

where $i=0,1,..., S$. Since $P_{i,0,1}s$ have been obtained, it is easy to find all $P_{i,j,1}$ by using the above equations. Thus, the general expression for the approximate state probabilities can be written as follows:

\begin{equation}
P_{i,j,1} = \begin{cases}
     \frac{(\frac{\eta}{\xi})^i}{i!\sum_{k=0}^{S}{\frac{(\frac{\eta}{\xi})^k}{k!}}}\Bigg [ \sum_{j=0}^{i} \frac{\rho^i}{j!} + \frac{(i^{L-i}\rho^{i+1})-\rho^{L+1}}{i!i^{L-1}(i-\rho)} \Bigg ] ^{-1} \frac{\rho^j}{j!} &    \\
       & \\
       where \ j = 0,1,2,иии ,i  & \\
       & \\
       \frac{(\frac{\eta}{\xi})^i}{i!\sum_{k=0}^{S}{\frac{(\frac{\eta}{\xi})^k}{k!}}}\Bigg [ \sum_{j=0}^{i} \frac{\rho^i}{j!} + \frac{(i^{L-i}\rho^{i+1})-\rho^{L+1}}{i!i^{L-1}(i-\rho)} \Bigg ] ^{-1} \frac{\rho^j}{j!i^{j-i}} &   \\
        & \\
       where \ j = i+1,i+2,иии ,L  & \\
    \end{cases}
\end{equation}

Hence, all $P_{i,j,1}$ can be calculated using equation 9 which are not exact. Please note that, the state probabilities obtained for $Plane_1$ are used to get faster computations and more accurate results. However, it is not possible to follow a similar approach for $Plane_0$. This is mainly because in $Plane_0$ the system does not serve.

\subsection{Balance Equations and Iterative Solution}

In this section, the main balance equations are derived for each plane individually in order to obtain all $P_{i,j,n}$. In addition, the balance equations obtained are 
also used to consider the transitions between these planes to take all possible transitions into account.
 
 \subsubsection{Balance Equations for $Plane_1$ and $Plane_0$}

In order to accumulate the effects of lateral and horizontal transitions together for the $Plane_1$, all possible balance equations can be derived from the Fig. 5. 
These transitions lead to obtain the following balance equations:

$\underline{i=1}$;\\
\\
$j=0$\\
\begin{equation}
P_{i,j,1} =  \frac{\mu P_{i,j+1,1}+(i+1)\xi P_{i+1,j,1}}{\eta + \lambda} 
\end{equation}
\\
$0 < j < L$\\

\begin{equation}
P_{i,j,1} =  \frac{\mu P_{i,j+1,1}+(i+1)\xi P_{i+1,j,1}+\lambda P_{i,j-1,1}}{\eta + \lambda+ \mu} 
\end{equation}

$j=L$\\

\begin{equation}
P_{i,j,1} =  \frac{(i+1)\xi P_{i+1,j,1}+\lambda P_{i,j-1,1}}{\eta + \mu}  
\end{equation}

$\underline{2 \leq i < S};$\\

$j=0$\\

\begin{equation}
P_{i,j,1} =  \frac{\mu P_{i,j+1,1}+(i+1)\xi P_{i+1,j,1}+\eta P_{i,j-1,1}}{ \lambda+ \eta + i\xi} 
\end{equation}

$1 \leq j < i$\\

\begin{equation}
P_{i,j,1} =  \frac{(j+1)\mu P_{i,j+1,1}+(i+1)\xi P_{i+1,j,1}+\lambda P_{i,j-1,1}\eta P_{i,j-1,1}}{ \lambda+ \eta + j\mu+ i\xi} 
\end{equation}

$1 \leq j < L$\\

\begin{equation}
P_{i,j,1} =  \frac{i\mu P_{i,j+1,1}+(i+1)\xi P_{i+1,j,1}}{ \lambda+ \eta + i\mu+ i\xi+ \lambda P_{i,j-1,1}+ \eta P_{i-1,j,1}} 
\end{equation}

$j = L$\\

\begin{equation}
P_{i,j,1} =  \frac{(i+1)\xi P_{i+1,j,1}+\lambda P_{i,j-1,1}+ \eta P_{i-1,j,1}}{ \eta + i\mu+ i\xi} 
\end{equation}

$\underline{i = S};$\\

$j=0$\\

\begin{equation}
P_{i,j,1} =  \frac{\mu P_{i,j+1,1}+\eta P_{i-1,j,1}}{ \lambda + i\xi} 
\end{equation}

$1 \leq j < i$\\

\begin{equation}
P_{i,j,1} =  \frac{(j+1)\mu P_{i,j+1,1}+\lambda P_{i,j-1,1}+ \eta P_{i-1,j,1}}{ \lambda + j\mu+ i\xi} 
\end{equation}

$i \leq j < L$\\

\begin{equation}
P_{i,j,1} =  \frac{i\mu P_{i,j+1,1}+\lambda P_{i,j-1,1}+ \eta P_{i-1,j,1}}{ \lambda + i\mu+ i\xi} 
\end{equation}

$j=L$\\

\begin{equation}
P_{i,j,1} =  \frac{\lambda P_{i,j-1,1}+\eta P_{i-1,j,1}}{i\mu+ i\xi} 
\end{equation}

On the other hand, the balance equations of the $Plane_0$ is also required to have an approximate solution of the proposed system. Therefore, similarly to $Plane_1$ the general balanced equations are produced considering the Fig. 4 and all possible balance equations are derived as follows:

$\underline{i = 0};$\\

$j=0$\\

\begin{equation}
P_{i,j,0} =  \frac{\xi P_{i+1,j,0}}{ \lambda} 
\end{equation}

$0 < j < L$\\

\begin{equation}
P_{i,j,0} =  \frac{\xi P_{i+1,j,0}+ \lambda P_{i,j-1,0}}{ \lambda} 
\end{equation}

$j=L$\\

\begin{equation}
P_{i,j,0} =  \xi P_{i+1,j,0}+ \lambda P_{i,j-1,0}
\end{equation}

$\underline{1 \leq i < S-1};$\\

$j=0$\\

\begin{equation}
P_{i,j,0} =  \frac{(i+1)\xi P_{i+1,j,0}}{ \lambda + i\xi } 
\end{equation}

$1 \leq j < and i \leq j < L$\\

\begin{equation}
P_{i,j,0} =  \frac{(i+1)\xi P_{i+1,j,0}+ \lambda P_{i,j-1,0}}{ \lambda + i\xi } 
\end{equation}

$j = L$\\

\begin{equation}
P_{i,j,0} =  \frac{(i+1)\xi P_{i+1,j,0}+ \lambda P_{i,j-1,0}}{i\xi} 
\end{equation}

$\underline{i=S-1};$\\

$1 \leq j < i and i \leq j < L$\\

\begin{equation}
P_{i,j,0} =  \frac{\lambda P_{i,j-1,0}}{ \lambda + i\xi } 
\end{equation}

$j = L$\\

\begin{equation}
P_{i,j,0} =  \frac{\lambda P_{i,j-1,0}}{ i\xi } 
\end{equation}

Therefore, the final values of $P_{i,j,0}$ can also be obtained. The general balance equations are derived for both planes separately. However, the relation between both plane and balance equations are also required for the proposed system in order to obtain more accurate and correct results which is given in the next section.

\subsubsection{Essential Balance Equations}

In order to obtain correct steady state probabilities of the proposed system the essential balance equations have to be considered. Thus, Fig. 7 indicates the three dimensional model considered which shows relation between two planes.

\begin{figure}[!htbp]
  \begin{center}
    \leavevmode
    \ifpdf
      \includegraphics[height=3in]{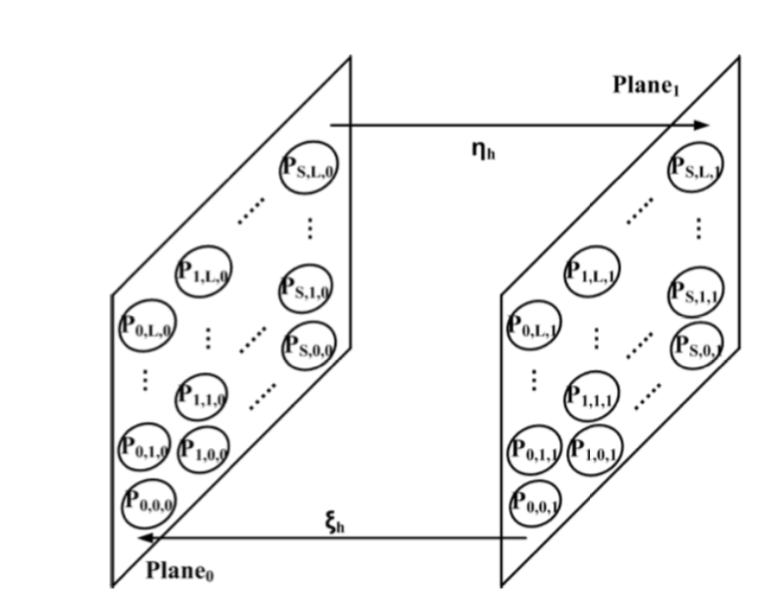}
              \fi    \caption{Three dimension (3D) feature of the proposed system
}
    \label{SM}
  \end{center}
\end{figure}

Thus, equation 29 gives the relation between two planes that can easily be obtained from the Fig. 7.

\begin{equation}
\eta hP_{i,j,0} =  \xi hP_{i,j,1} 
\end{equation}

Hence, the relation between two planes with essential balance equations for each state can be obtained by using given relation in equation 29. The essential balance equations obtained for $Plane_1$ considering the effects of $Plane_0$ are as follows:

$\underline{i=1};$\\

$j=0$\\

\begin{equation}
P_{i,j,1} =  \frac{\mu P_{i,j+1,1}+ (i+1)\xi P_{i+1,j,1}+ \eta hP_{i-1,j,0}}{ \lambda + \eta+ \xi h } 
\end{equation}

$1 < j < L$\\

\begin{equation}
P_{i,j,1} =  \frac{\mu P_{i,j+1,1}+ (i+1)\xi P_{i+1,j,1}+ \lambda P_{i,j-1,1} + \eta hP_{i-1,j,0}}{ \lambda + \eta+ \mu + \xi h } 
\end{equation}

$j = L$\\

\begin{equation}
P_{i,j,1} =  \frac{(i+1)\xi P_{i+1,j,1}+ \lambda P_{i,j-1,1} + \eta hP_{i-1,j,0}}{ \eta+ \mu + \xi h } 
\end{equation}

$\underline{2 \leq i < S;}$\\

$j=0$\\

\begin{equation}
P_{i,j,1} =  \frac{\mu P_{i,j+1,1}+ (i+1)\xi P_{i+1,j,1}+ \eta P_{i-1,j,1} + \eta hP_{i-1,j,0}}{ \lambda + \eta+ i\xi + \xi h } 
\end{equation}

$1 \leq j < i $\\
\begin{equation}
P_{i,j,1} =  \frac{(j+1)\mu P_{i,j+1,1}+ (i+1)\xi P_{i+1,j,1}+ \lambda P_{i,j-1,1}}{ \lambda + \eta+ j\mu+ i\xi + \xi h } +  \\ 
 \frac{\lambda P_{i,j-1,1}+\eta P_{i-1,j,1}+\eta hP_{i-1,j,0}}{ \lambda + \eta+ j\mu+ i\xi + \xi h }
\end{equation}

$i \leq j < L $\\
\begin{equation}
P_{i,j,1} =  \frac{i\mu P_{i,j+1,1}+ (i+1)\xi P_{i+1,j,1}+ \lambda P_{i,j-1,1}}{ \lambda + \eta+ i\mu+ i\xi + \xi h } +  \\ 
 \frac{\eta P_{i-,j,1}+\eta hP_{i-1,j,0}}{ \lambda + \eta+ i\mu+ i\xi + \xi h }
\end{equation}

$j = L $\\
\begin{equation}
P_{i,j,1} =  \frac{(i+1)\xi P_{i+1,j,1}+ \lambda P_{i,j-1,1}+ \eta P_{i-1,j,1}+\eta hP_{i-1,j,0}}{ \eta+ i\mu+ j\xi + \xi h } 
\end{equation}

$\underline{ i = S};$\\

$j=0$\\

\begin{equation}
P_{i,j,1} =  \frac{\mu P_{i,j+1,1}+ \eta P_{i-1,0,1}+ \eta hP_{i-1,j,0}}{ \lambda+ i\xi + \xi h } 
\end{equation}

$1 \leq j < i$\\

\begin{equation}
P_{i,j,1} =  \frac{(j+1)\mu P_{i,j+1,1}+ \lambda P_{i,j-1,1}\eta P_{i-1,j,1}+ \eta hP_{i-1,j,0}}{ \lambda+ j\mu+ i\xi + \xi h } 
\end{equation}

$i \leq j < L $\\

\begin{equation}
P_{i,j,1} =  \frac{(i)\mu P_{i,j+1,1}+ \lambda P_{i,j-1,1}\eta P_{i-1,j,1}+ \eta hP_{i-1,j,0}}{ \lambda+ i\mu+ i\xi + \xi h } 
\end{equation}

$j = L $\\

\begin{equation}
P_{i,j,1} =  \frac{\lambda P_{i,j-1,1}\eta P_{i-1,j,1}+ \eta hP_{i-1,j,0}}{i\mu+ i\xi + \xi h } 
\end{equation}

In addition, the essential balance equations obtained for $Plane_0$ considering the effects of $Plane_1$ are as follows:

$\underline{i = 0};$\\

$j=0$

\begin{equation}
P_{i,j,0} =  \frac{\xi P_{i+1,j,0}+\xi hP_{i+1,j,1}}{\lambda + \xi h } 
\end{equation}

$0 < j < L$

\begin{equation}
P_{i,j,0} =  \frac{\xi P_{i+1,j,0}+ \lambda P_{i,j-1,0}+ \xi hP_{i+1,j,1}}{\lambda + \xi h } 
\end{equation}

$j = L$

\begin{equation}
P_{i,j,0} =  \frac{\xi P_{i+1,j,0}+ \lambda P_{i,j-1,0}+ \xi hP_{i+1,j,1}}{\xi h } 
\end{equation}

$\underline{1 \leq i < S-1};$\\

$j=0$\\

\begin{equation}
P_{i,j,0} =  \frac{(i+1)\xi P_{i+1,j,0} + \xi hP_{i+1,j,1}}{\lambda + i\xi + \xi h } 
\end{equation}

$0 < j < L$\\

\begin{equation}
P_{i,j,0} =  \frac{(i+1)\xi P_{i+1,j,0} + \lambda P_{i,j-1,0} + \xi hP_{i+1,j,1}}{\lambda + i\xi + \xi h } 
\end{equation}

$j = L$\\

\begin{equation}
P_{i,j,0} =  \frac{(i+1)\xi P_{i+1,j,0} + \lambda P_{i,j-1,0} + \xi hP_{i+1,j,1}}{i\xi + \xi h } 
\end{equation}

$\underline{i=S-1};$\\

$j=0$\\ 
\begin{equation}
P_{i,j,0} =  \frac{\xi hP_{i+1,j,1}}{\lambda + i\xi + \eta h } 
\end{equation}

$0 < j < L$\\
\begin{equation}
P_{i,j,0} =  \frac{\lambda P_{i,j-1,0}+\xi hP_{i+1,j,1}}{\lambda + i\xi + \eta h } 
\end{equation}

$ j = L$\\
\begin{equation}
P_{i,j,0} =  \frac{\lambda P_{i,j-1,0}+\xi hP_{i+1,j,1}}{(i-1)\xi + \eta h } 
\end{equation}

Then, the iterative procedure can be applied in order to accurately calculate the $P_{i,j,n}$. The iterative procedure can be given as follows:

\begin{enumerate}
\item{\textbf{Input:} Define the input parameters; $S, L, \lambda, \mu h, \mu, \eta h, \eta, \xi h, \xi$, the maximum number of iterations i.e, $2^{18}$ and the converge parameter $\triangle$}
\item{Equation 9 calculates the approximate steady state probabilities, $P_{i,j,1}$ of the $Plane_1$ to get faster computations and more accurate results. Equations 10-20 and 21-28 are then used to calculate rough steady state probabilities of the system considered for both planes. These equations and computations are used to have a faster convergence when the iterative procedure of essential balance equations is employed. For instance, $MQL_{converge}$ is obtained using mathematical relations in application as follows:
Set the initial value of $MQL_{converge}$ = 0.0
\begin{verbatim}
for (i=0; i \leq S; i++) do
 	for (j=0; j \leq L; j++) do
	$MQL_{converge} = MQL_{converge} +j*P_{i,j,n} $
	end for
end for
\end{verbatim}
}
\item{The balance equations given in equations 29-49 are used to calculate the correct steady state probabilities, $P_{i,j,n}$.}

\item{The sum of all  $P_{i,j,n}$. can be determined by the help of the normalizing condition. Steps 3 and 4 are repeated until the normalization condition is satisfied. In other words, the sum of probabilities sufficiently converges to one. Thus, a performability measures, MQL, is chosen to check the accuracy by the convergence parameter $\triangle$, where $\triangle=0.001$ is taken in this paper.
\begin{verbatim}
for (i=0; i \leq S; i++) do
 	for (j=0; j \leq L; j++) do
	$MQL_{approx} = MQL_{approx} +j*P_{i,j,n} $
	end for
end for
\end{verbatim}
Thus, the iteration will be ended in case $\| MQL_{approx} - MQL_{converge}\| \leq \triangle$. Otherwise, the iterative procedure will assign the recent values of performability measures to old values i.e, $MQL_{converge} = MQL_{approx}$ and continue from the step 3 separately for both planes.
}
\item{Hence, the performability measures such as mean queue length (MQL), $MQL = MQL_0 + MQL_1$, throughput (THRP), $THRP = THRP_0 + THRP_1$ and mean response time (MRT), $MRT = MRT_0 + MRT_1$ can be computed using the most recent steady state probabilities $P_{i,j,n}$ as: \\

$MQL = \sum_{i=1}^{S-1}\sum_{j=0}^{L} P_{i,j,0} + \sum_{i=1}^{S}i \sum_{j=0}^{L} P_{i,j,1} $ \\

$THRP = \sum_{i=1}^{S-1}\sum_{j=0}^{L} j\mu P_{i,j,0} + \sum_{i=1}^{S}\sum_{j=0}^{L} j\mu P_{i,j,1} $ \\

$MRT = \frac{MQL}{THRP}$
}

\end{enumerate}

\section{Numerical Results and Discussions}
In this section, the results are presented for the performability evaluation of the large-scale Beowulf clusters considering a head node and a large number of computing nodes. Numerical results are presented for MQL, THRP, and MRT in order to show the capability of the proposed analytical model and the solution approach. Numerical results obtained clearly show that, the proposed model and an approximate solution approach can easily handle large number of nodes in Beowulf clusters without a state explosion problem. In addition, numerical results obtained from the proposed model are validated by DES. The CPU times of the proposed solution approach and DES are also compared in order to show the efficacy of the proposed model and an approximate solution approach.

The mean arrival and service rates are mainly application dependent. The assumptions and parameters used in [3,7,9,11-13,32] are also employed in this paper for consistency. However, the approach presented in this paper is flexible and can be adopted to similar applications. For instance, the average failure and repair rates of the computing nodes considered are chosen from studies [3], and [7] considering the availability of Beowulf clusters as well as cloud computing systems for consistency. Therefore, time between failures for head node and computing nodes can be taken as $250 (\xi=\xi h= 0.004/h), 500 (\xi = \xi h= 0.002/h)$, and $1000 (\xi =\xi h= 0.001/h)$ hours. The parameters are taken for many computations as $\mu=0.25 requests/sec, \eta=\eta h=0.5/hr, \xi= \xi h=0.001/hr$ and the $\lambda$ rate per user varies from 10 requests per second unless stated otherwise. MQL, THRP, and MRT results are presented as a function of $\lambda$ for proposed analytical model and DES in Figs. 8-10, respectively with different computing node failure rates. A system with $S=500$, and $L=1000$ is considered.

\begin{figure}[!htbp]
  \begin{center}
    \leavevmode
    \ifpdf
      \includegraphics[height=3in]{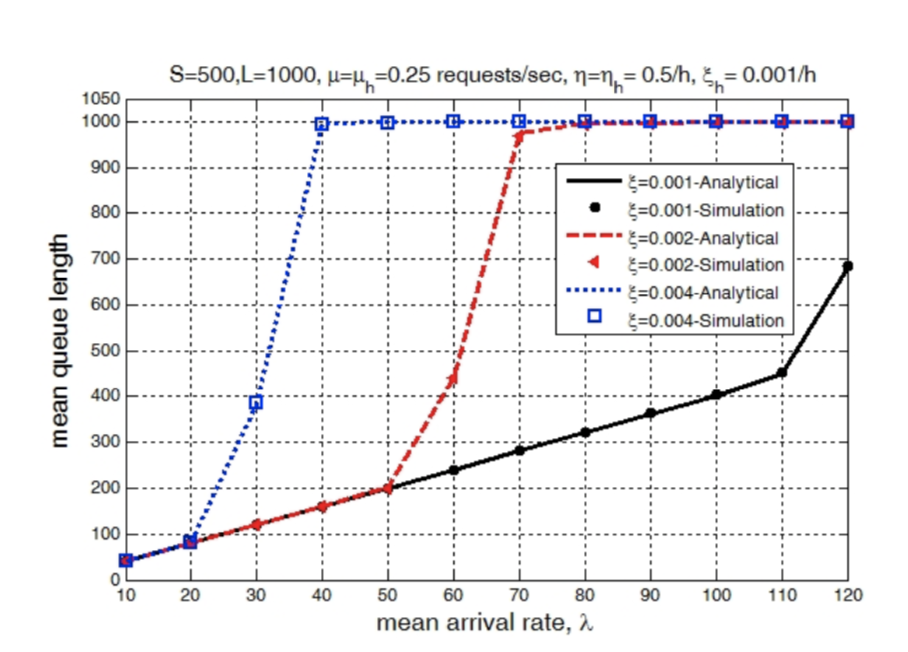}
              \fi    \caption{MQL vs $\lambda$ for S=500 and L=1000 with different $\xi$.}
    \label{SM}
  \end{center}
\end{figure}

\begin{figure}[!htbp]
  \begin{center}
    \leavevmode
    \ifpdf
      \includegraphics[height=3in]{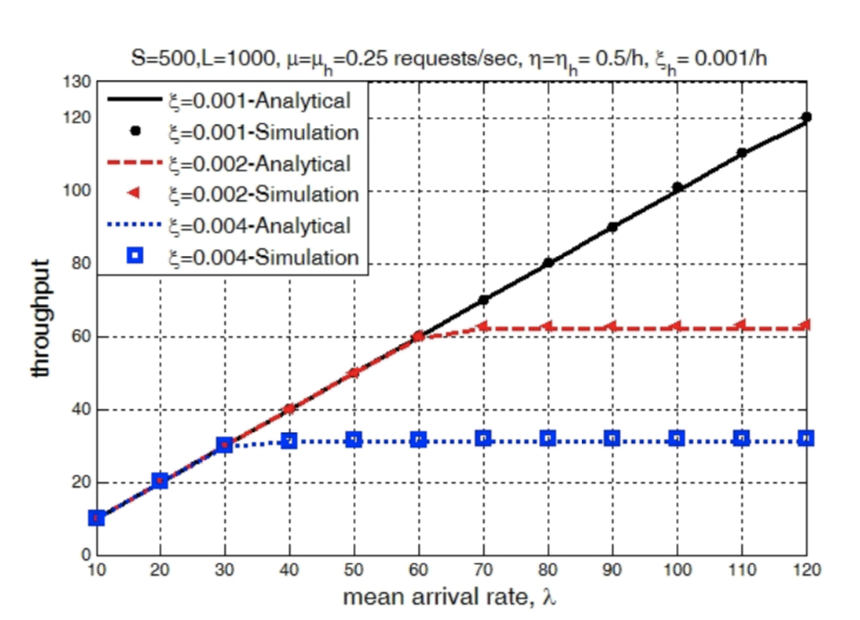}
              \fi    \caption{THRP vs $\lambda$ for S=500 and L=1000 with different $\xi$}
    \label{SM}
  \end{center}
\end{figure}

The figures clearly show that the effects of computing nodes failure on the QoS of the system is quite significant in large scale fault tolerant Beowulf clusters. The system can be full quickly when a higher value of $\xi$ is considered. However, the considered Beowulf cluster can serve the arriving requests due to the large number of computing nodes for lower values of $\xi$. In other words, increasing the time between failures of computing nodes decreases the MQL as shown in Fig. 8. For instance, in Fig. 8, the MQL value is 280.924 when $\xi=0.001$ and $\lambda = 70$. However, increasing the failure rate of computing nodes to 0.002 and 0.004 when $\lambda = 70$, the MQL values increase to 976.147 and 999.188, respectively. In addition, the average number of requests in the system becomes same as the maximum capacity of the system L when the failure rate increases. It can be clearly observed in Fig. 9 that THRP of the system increases as $\lambda$ increases, however, the THRP saturates after some point depending on as well as the failure rates. This is because of the system cannot serve the requests efficiently and incoming requests start to queue up in the system to be served especially for the loaded systems. Higher THRP values are obtained for the systems with lower failure rates due to the average value of operative computing nodes. On the other hand, similar behaviour is observed for MRT in Fig. 10. The MRT increases when the failure rate of computing node increases as expected.

On the other hand, the effects of the failure and the repair rates of head node are given in Figs. 11 and 12, respectively. Figure 11 shows the THRP results as a function of arrival rate for different head node failure rates. The THRP results, decrease when $\xi h=0.01$. This is due to the frequent failing of the head node since the computing nodes directly depend on the head node. Even though a good repair facility is provided to the system $(\eta  = \eta h=0.5/h)$, the QoS degrades due to the failure rate of the head node. In addition, the MQL results as a function of arrival rate for different repair rates of the head node is given in Fig. 12.

\begin{figure}[!htbp]
  \begin{center}
    \leavevmode
    \ifpdf
      \includegraphics[height=3in]{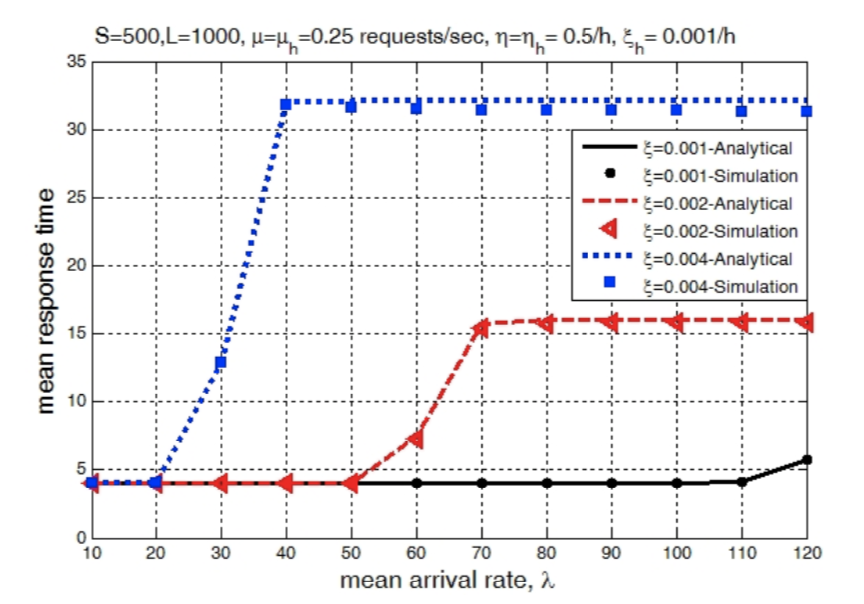}
              \fi    \caption{MRT vs $\lambda$ for S=500 and L=1000 with different $\xi$.}
    \label{SM}
  \end{center}
\end{figure}

\begin{figure}[!htbp]
  \begin{center}
    \leavevmode
    \ifpdf
      \includegraphics[height=3in]{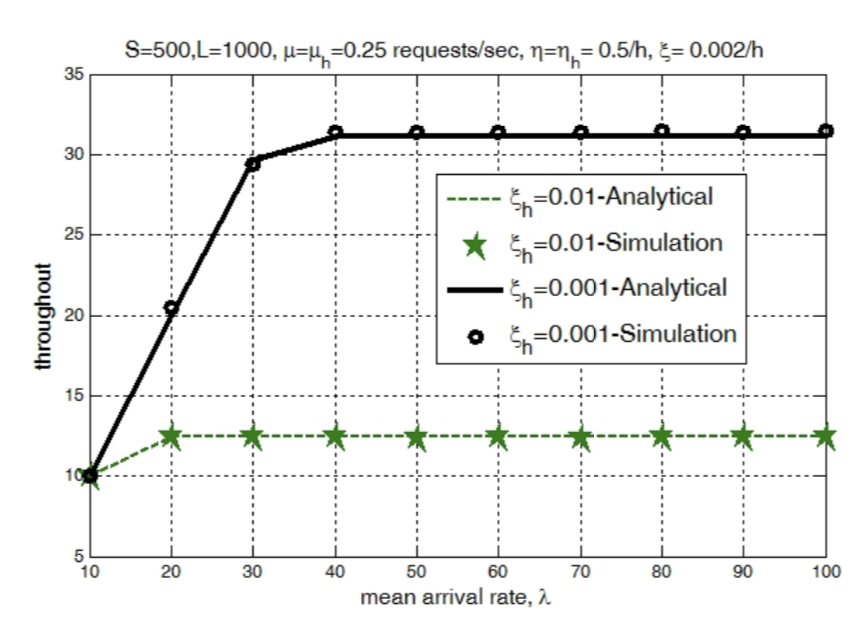}
              \fi    \caption{THRP vs $\lambda$ with different $\xi h$.}
    \label{SM}
  \end{center}
\end{figure}

\begin{figure}[!htbp]
  \begin{center}
    \leavevmode
    \ifpdf
      \includegraphics[height=3in]{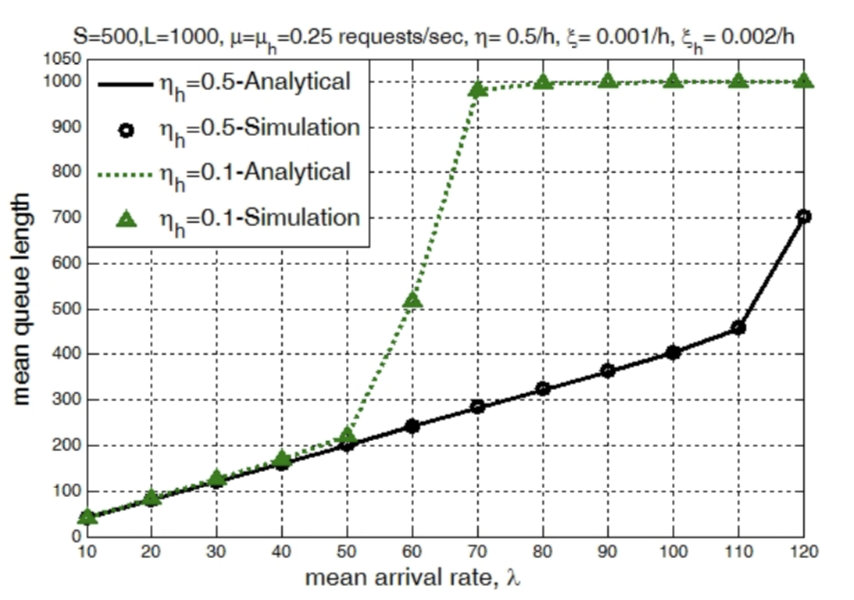}
              \fi    \caption{MQL vs $\lambda$ with different $\eta h$.}
    \label{SM}
  \end{center}
\end{figure}

The significance of the head node repair rates is clearly shown in the figure in terms of the queue length. In other words, when the system has a light traffic $(\lambda = 70)$, the MQL is 281.849. However, for the same situation the MQL is almost full (999.189) due to the frequent failing of the head node. In the case of the frequent failure of the head node, the rest of the computing nodes are no longer able to serve. Hence, the MQL of the system increases rapidly. In addition THRP of the system decreases because the system does not serve. If a head node is unable to serve (fails), the system stops serving the requests. The head node may regain connectivity and rejoin the system after some time. This situation is clearly shown in Fig. 12. Hence, as a summary, the repair and failure rates of the head node has a significant affect on the such system performance. On the other hand, Fig. 13 shows the MQL results as a function of arrival rate for different number of computing nodes. The various numbers of computing nodes are taken from [16] (S=32,64,128,256 and 372) in Fig. 13 where the researchers currently use them in their laboratories for different purposes. As can be seen that the proposed model and solution can easily handle the large amount of computing nodes up to several hundreds with availability issues. Fig. 14 shows the MQL results as a function of arrival rate for different queue capacities. As can be clearly seen from the figure that queue capacity is the limiting factor of large scale Beowulf clusters.

\begin{figure}[!htbp]
  \begin{center}
    \leavevmode
    \ifpdf
      \includegraphics[height=3in]{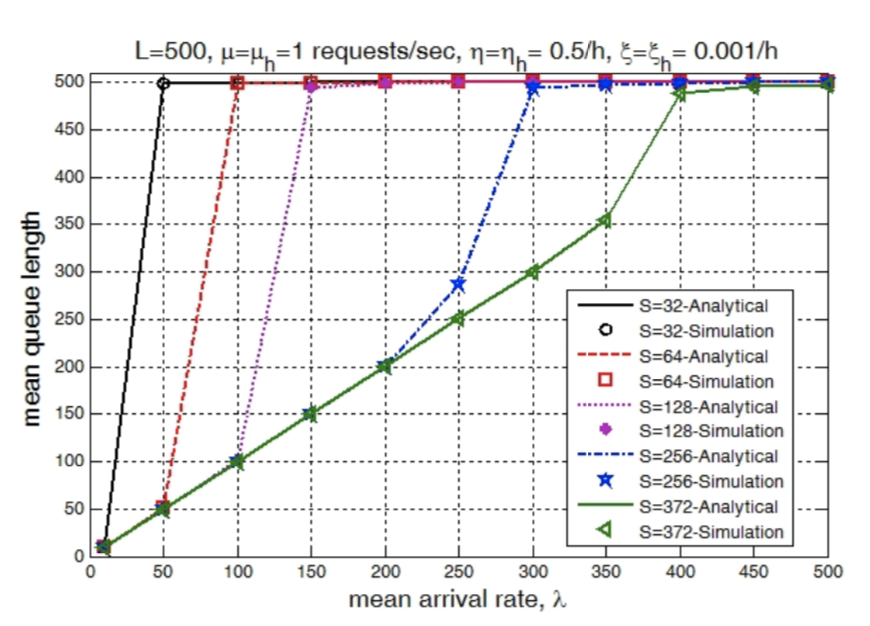}
              \fi    \caption{MQL vs $\lambda$ with different number of nodes.}
    \label{SM}
  \end{center}
\end{figure}

\begin{figure}[!htbp]
  \begin{center}
    \leavevmode
    \ifpdf
      \includegraphics[height=3in]{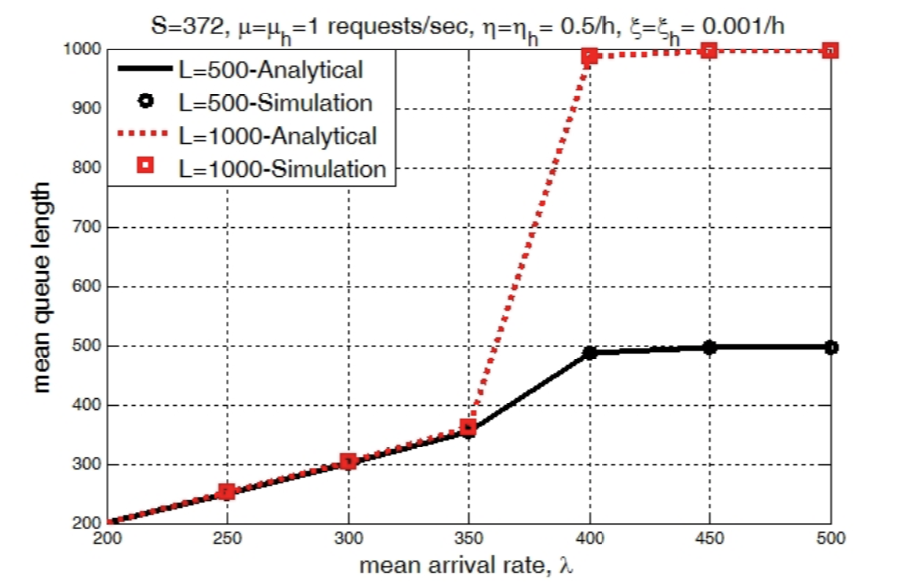}
              \fi    \caption{MQL vs $\lambda$ with different queue capacities.}
    \label{SM}
  \end{center}
\end{figure}

A comparative study is further performed in order to show a certain degree of accuracy of the proposed solution approach and DES. Tables 1 and 2 present MQL, THRP and MRT results comparatively with the simulation results for S=500, L=1000 and S=1000, L=2000, respectively.
The discrepancies of the proposed analytical model and DES are also presented in all the tables. The maximum discrepancies for MQL, THRP and MRT are less than 1.149$\%$, 3.82$\%$, and 3.76$\%$, respectively, for both tables which is well within the 5$\%$ confidence interval of the simulation. The DES is mainly used for the validation purposes, however it can also be used for the performance evaluation of such systems. Because it simulates the actual scenario rather than the Markov models presented in this paper.

The DES model is implemented in C++ language and adopted for the scenario considered [33]. In addition, the CPU times of the analytical approach and DES for the computations are also presented comparatively in Tables 3 and 4. All of the numerical results presented are obtained using workstations with Intel(R) Core(TM) i7-363QM CPU @ 2.40GHz, 16GB RAM, and 64-bit operating system.

The proposed 3D analytical model uses an iterative approach to obtain steady state probabilities based on $(S)x(L+1)$ number of equations for both $Plane_0$ and $Plane_1$. For instance, the number of states is considered in Table 4 is (1000 x 2001) x (1000 x 2001). Thus, the processing times of analytical models are also presented with the processing times of the simulation for comparison. Tables 3 and 4 show the CPU times of systems with S=500, L=1000, and S=1000, L=2000, respectively. The computational efficiency of the proposed solution approach with the DES is clearly given in both tables in terms of CPU times. For example, in Table 4 the maximum CPU time for simulation is 80878.871 seconds (22.46 hours) for S=1000, L=2000 whereas the maximum CPU time of the analytical approach is less than 5hours for an extreme case. Thus, the proposed analytical model and an approximate solution approach are efficient in performability evaluation of large-scale Beowulf clusters.

\begin{figure}[!htbp]
  \begin{center}
    \leavevmode
    \ifpdf
      \includegraphics[height=4.3in]{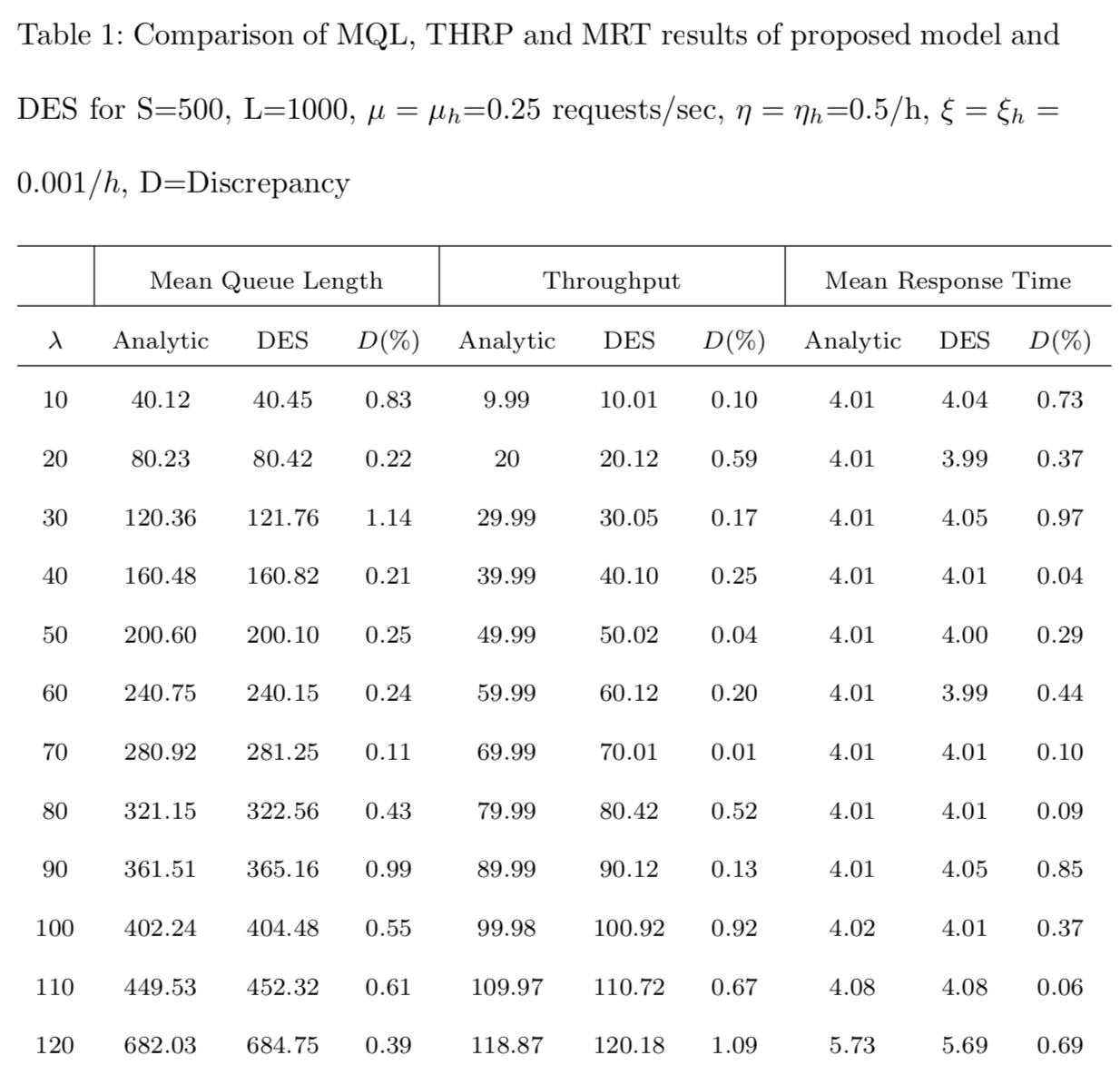}
              \fi  
    \label{SM}
  \end{center}
\end{figure}
 
\begin{figure}[!htbp]
  \begin{center}
    \leavevmode
    \ifpdf
      \includegraphics[height=2in]{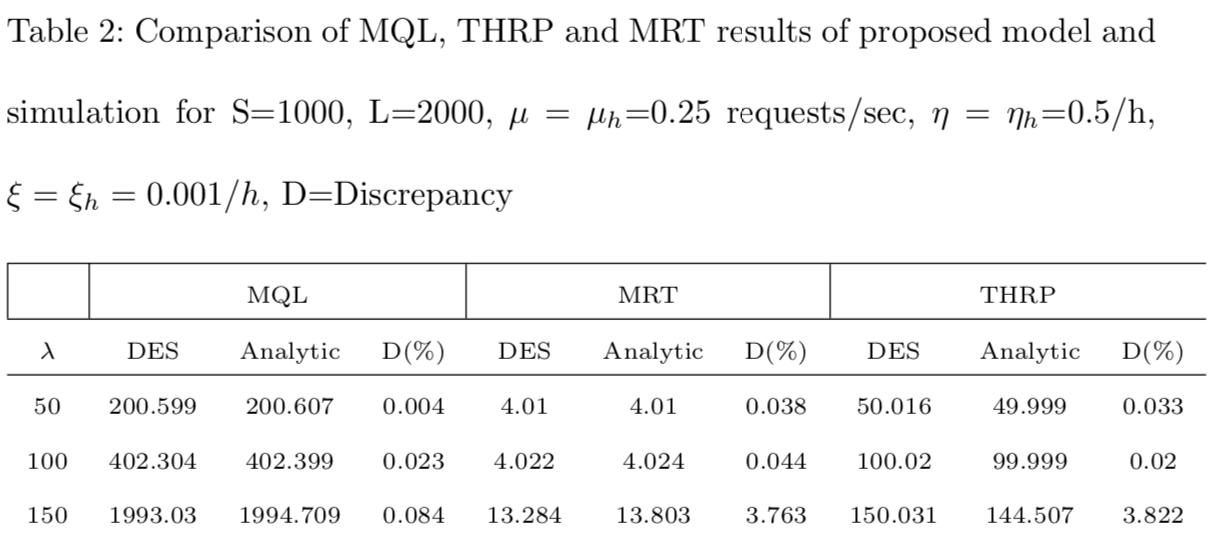}
              \fi  
    \label{SM}
  \end{center}
\end{figure}

\begin{figure}[!htbp]
  \begin{center}
    \leavevmode
    \ifpdf
      \includegraphics[height=4.3in]{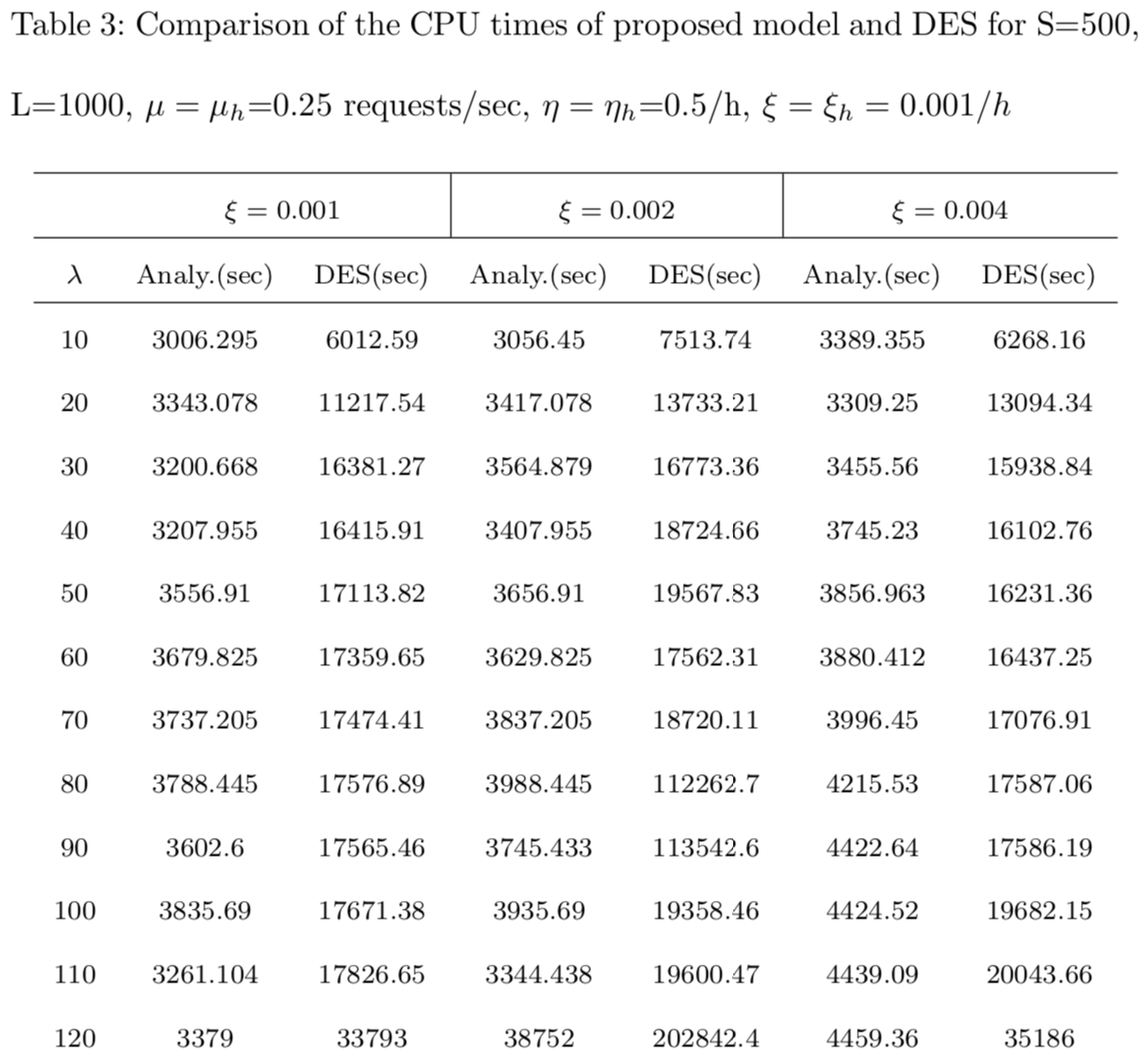}
              \fi  
    \label{SM}
  \end{center}
\end{figure}

\begin{figure}[!htbp]
  \begin{center}
    \leavevmode
    \ifpdf
      \includegraphics[height=1.7in]{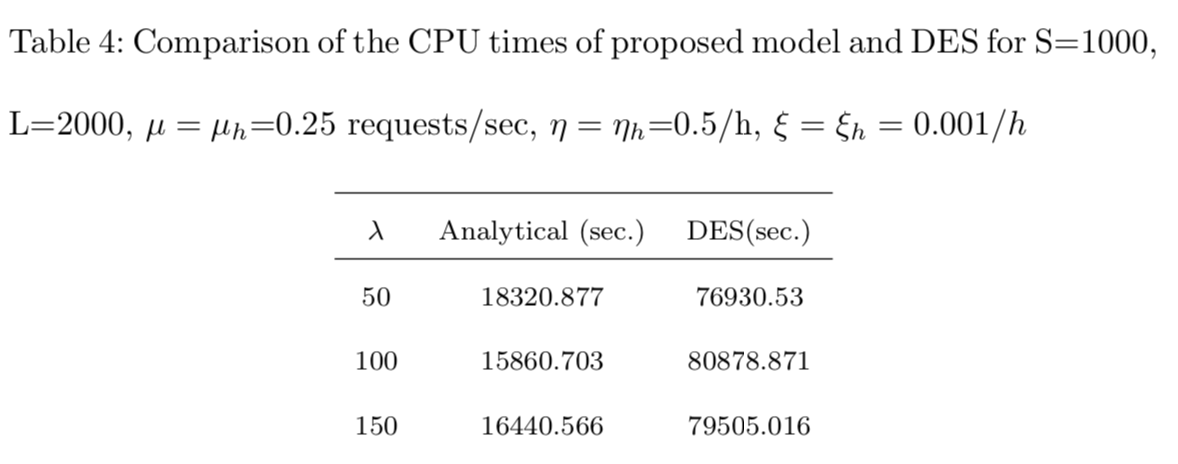}
              \fi  
    \label{SM}
  \end{center}
\end{figure}

\section{Conclusion}

This paper proposed an analytical modelling approach and an approximate solution approach to obtain QoS measurements for large-scale Beowulf clusters without a state explosion problem. In order to obtain more realistic QoS measurements, availability issues are considered together with performance modelling. The proposed analytical modelling, solution, and the analysis is useful for achieving better performance in such systems. The system is modelled as a three-dimensional Continuous Time Markov chain to determine the state probabilities. The proposed model can be used to analyze QoS measures such as mean queue length (MQL), throughput (THRP) and mean response time (MRT).

The method used is novel and flexible where can be extended to the case of many other practical, fault-tolerant multi-server farms. In addition, the numerical results presented show that the proposed solution approach can handle up to several million states of the model presented. A number of states

can be increased using the proposed model without having a state explosion problem depending on the system. The comparative results presented in this paper show that the discrepancy between the simulation and the analytical model presented is less than 5$\%$ for all cases. In terms of efficacy, the computation time of the proposed model is significantly shorter than the simulation especially for loaded and large-scale cases.

The main demerit of this method is the large number equations. In the proposed model, the balance equations depend on each other and chained together for obtaining approximate steady state probabilities. An iterative technique has been used to solve the steady state probabilities. Also, this technique increases the computation times for approximate results. Furthermore, increasing the queue capacity, the number of computing nodes, or mean arrival rate forces a significant increase in computation times. Therefore, it is essential that programming techniques are effectively used to further reduce computation times. This work is still in progress. Although the speed is an issue, the proposed method is superior to simulation under all circumstances. Results show that, proposed method works with large queue capacities and large numbers of computing nodes effectively giving accurate results.

\end{document}